\documentclass[a4paper,11pt]{article}

\usepackage{jheppub1} 

\usepackage[latin3]{inputenc}
\usepackage[makeroom]{cancel}
\usepackage{graphicx}
\usepackage{amsmath}
\usepackage{amsfonts}
\usepackage{amssymb}
\usepackage{color}
\usepackage[export]{adjustbox}
\usepackage{graphicx}
\usepackage{dcolumn}
\usepackage{bm}
\usepackage{simplewick}
\usepackage{array}
\usepackage{appendix}
\usepackage{relsize}
\usepackage{subcaption}

\newcommand{\be}{\begin{equation}}
	\newcommand{\ee}{\end{equation}}
\newcommand{\beq}{\begin{equation}}
	\newcommand{\eeq}{\end{equation}}
\newcommand{\bea}{\begin{eqnarray}}
	\newcommand{\eea}{\end{eqnarray}}

\title{\boldmath Thermodynamics of dynamical wormholes}

\author{Mudassar Rehman and}
\author{Khalid Saifullah}

\affiliation {Department of Mathematics, Quaid-i-Azam University, Islamabad, Pakistan}

\emailAdd{mrehman@math.qau.edu.pk}
\emailAdd{ksaifullah@fas.harvard.edu}
\abstract{We study thermodynamics of dynamical traversable wormholes. These wormholes are investigated in the background of different cosmological  models, with and without the cosmological constant, and which include the power-law and exponential cosmologies also. We work out the generalized surface gravity for wormholes of different shapes. The surface gravity is evaluated at the trapping horizon and the unified first law of thermodynamics is set up. The thermodynamic stability of these wormholes has also been investigated. Some cases of asymptotically flat, de Sitter and anti-de Sitter wormholes have been considered as well. Our results generalize those that exist for static Morris-Thorne wormholes.
\vspace{85 mm}
}

\notoc

\begin{document}
\maketitle


\section{Introduction}

\label{sec:intro}

The idea of wormholes is not new and it was discussed in early 20th
century by some authors including Flamm \cite{R1}, Weyl \cite{R2} and Einstein and Rosen \cite{R3} but the name \emph{wormhole} was first used
by Misner and Wheeler \cite{R4}. Recently a considerable interest in wormhole physics has been seen in two directions: one with the Euclidean signature metrics and the other with the Lorentzian wormholes \cite{R4A, R4B}. Lorentzian wormholes that are both stable and
traversable were first investigated by Morris and Thorne \cite{R5} in 1988.
Wormholes provide shortcuts to go from one universe to the other or
from one part to the other part of the same universe. For a
wormhole to be traversable it must not have an event horizon. This requires that the spacetime contains some unusual or exotic matter. This means that the matter has very strong negative pressure and even the energy density is negative according to the static observer.
Here, in this paper, the term wormhole would mean a traversable
wormhole. Another interesting property of a wormhole is that it can be converted into
time machine if one of its mouth is moved relative to the other \cite{R6}.

The standard cosmology reveals the fact that the total dominating energy density of the universe is in the form of dark matter and dark energy. The latter is considered to be uniformly distributed all over the universe and associated with a negative pressure and accelerates the expansion of the universe. Dark energy explained entirely on the basis of the cosmological constant is fully consistent with existing observational data. Another candidate is the phantom matter whose energy density increases with the expansion of the universe and which is associated with negative pressure \cite{R6A, R7, R8}. This matter violates the null energy condition and it could be the type of matter which supports wormhole
structure \cite{R9, R10}. This provides evidence that wormholes could
exist in the real universe and that they are not just a mathematical
toy spacetime model. Now, exotic matter is considered to be a time-reversed version of
ordinary matter, therefore, one may think of wormhole also to be a time-reversed
version of black hole if both show similar thermodynamic behavior. These kinds of analyses will improve
the physical status of wormholes greatly \cite{R11, R12}.

The main aim of this paper is to investigate dynamical wormholes with particular refernce to their thermodynamic properties at trapping horizons which are the hypersurfaces foliated by
marginal surfaces. The need and significance of characterizing black holes by using local
considerations has been stressed by Hayward \cite{R13, R14, R15, R16}. Black holes are described
by the presence of event horizons, which is the global property and hence
cannot be located by observers. Now, trapping horizon is a pure local
concept, and in this way the thermodynamic properties of
spherically symmetric dynamical black holes were studied using local considerations. For wormholes, we will employ the definition of surface gravity \cite{R23} where we will use trapping horizon
instead of Killing horizon and Kodama vector will play the role of Killing
vector. The thermodynamic properties can also be studied for a wormhole by virtue of the
presence of trapping horizon, and the results analogous to those of a black hole can be obtained \cite{R17}. We investigate wormholes of different shapes for their thermodynamic properties within the framework of various cosmological models. These include asymptotically flat, de Sitter and anti-de Sitter wormholes as well.

In this paper Section 2 describes the trapping horizon of a spherical
symmetric dynamical wormholes. In Section 3 we find the generalized surface
gravity for these wormholes on a trapping horizon. Sections 4, 5 and 6 deal with thermodynamics of wormholes with different shape functions within the framework of different cosmological models. The unified first
law of wormhole thermodynamics is described in Section 7. Section 8 deals with the thermodynamic stability of wormholes.
In Section 9 we derive the expression for the surface gravity following the same approach as in Section 3 but now using the areal radius coordinates. We conclude our work in the last section.

\section{Trapping horizon}

The Hayward formalism uses local quantities to define the properties of real
black holes from which one obtains the same results that are yielded by the global considerations
in the static case using event horizons and when there is vacuum. It is interesting to note that wormhole thermodynamic properties are similar to those
found in black holes when we use local physically
relevant quantities. Since event horizon is not present in a traversable wormhole so we
use the trapping horizon. Now, the Schwarzschild black hole is the static
vacuum  solution that has a wormhole extension called the Einstein-Rosen
bridge. But it is not traversable as it contains an event horizon. Here we
consider a dynamical wormhole in a cosmological background, which is a generalization of the Morris-Thorne wormhole to a time dependent background \cite{R18},
\begin{equation}  \label{1a1}
ds^{2}=-e^{2\Phi (t,r)}dt^{2}+a^{2}(t)\left[\frac{dr^{2}}{1-\frac{b(r)}{r}}+r^{2}d\Omega^{2}\right],
\end{equation}
in coordinates $(t,r,\theta,\phi)$ where $d\Omega^{2}=d\theta ^{2}+\sin
^{2}\theta d\phi ^{2}$. The radial coordinate $r$ ranges in $[r_{0}, \infty]$. Here the minimum radius $r=r_{0}$ corresponds to the throat of the wormhole which connects two regions, each region is $r_{0}<r<r_{a}$ where $r_{a}$ corresponds to the radius of the wormhole mouth. At $r\rightarrow \infty$ this metric becomes flat, $a(t)$ is the dimensionless parameter called the scaling factor of the universe. It tells us how our universe is expanding. It is known that the expansion rate of our universe is increasing with time which implies $\ddot{a}(t)>0$ or $\dot{a}(t)$ is an increasing function of time (here over dot represents the time derivative). $\Phi(t,r)$ is the redshift function as it corresponds to the gravitational redshift. This function should be finite everywhere in order to prevent the existence of an event horizon which is the necessary requirement for a wormhole to be traversable and when $r\rightarrow \infty$ this redshift function should vanish. Here $b(r)$ is the shape function which describes the shape of a wormhole as can be seen from the embedding space in coordinates $(Z, r, \phi )$, where the 2-surface
 \begin{equation}\label{1aa}
   Z(r)=\pm \int \left(\frac{r}{b(r)}-1\right)^{-1/2}dr
 \end{equation}
 has the same geometry as the 2-surface $\theta =\pi /2$ and $t= \> constant$ in metric (\ref{1a1}). The function $Z(r)$ is called the embedding function. The graph of Eq. (\ref{1aa}), when revolved around the axis of rotation, the $Z$-axis, gives the shape of the wormhole \cite{R25}. At the wormhole throat a coordinate singularity $b(r_{0})=r_{0}$ occurs and $b(r)<r$ for $r>r_{0}$.  This condition ensures the finiteness of the proper radial distance defined by
 \begin{equation}\label{}
   l(r)=\pm \int _{r_{0}}^{r} \frac{dr}{\sqrt{1-\frac{b(r)}{r}}}
 \end{equation}
 where $\pm$ refers to the two asymptotically flat regions that are connected through the wormhole throat.
 The flaring out condition for wormholes requires that $b^{\prime}<b(r)/r$ at or near the throat which results in violating the null energy condition \cite{R5, R19, R20}. These are the conditions on $\Phi(t,r)$ and $b(r)$ which provide a stable wormhole solution. The stability of some static wormholes has been discussed in the literature \cite{LL, GGS, GGS1, RT}. It is clear that when $\Phi (t,r)$ and $b(r)$ tend to zero then the metric (\ref{1a1}) becomes the flat Friedmann-Robertson-Walker (FRW) metric, and Morris-Thorne metric is recovered when $\Phi (t,r)=\Phi(r)$ and $a(t)\rightarrow 1$. Moreover there are conditions that must be satisfied and the forces felt by the observer in the wormhole during his hypothetical travel which has been discussed in detail in Ref. \cite{R5}.  Here in this paper we take $\Phi (t,r)=0$ so that the wormhole metric (\ref{1a1}) takes the form
\begin{equation}  \label{1}
ds^{2}=-dt^{2}+a^{2}(t)\left[\frac{dr^{2}}{1-\frac{b(r)}{r}}+r^{2}d\Omega^{2}\right].
\end{equation}

Now for the energy-momentum tensor we take the perfect fluid which is completely described by its energy density and isotropic pressure \cite{R18}, with components
\begin{equation}  \label{1a}
T^{t}_{t}=-\rho (t,r), \> T^{r}_{r}= \> p_{r}(t,r), \> T^{\theta}_{\theta}= \> T^{\phi}_{\phi}= \> p_{t}(t,r),
\end{equation}
where $\rho (t,r)$, $p_{r}(t,r)$ and $p_{t}(t,r)$ are, respectively, the energy density, radial pressure and tangential pressure. For isotropic pressure  $p_{r}(t,r) = p_{t}(t,r)$, otherwise the pressure will be anisotropic.

The null coordinates for the above metric (\ref{1}) are given
by
\begin{equation}  \label{2}
x^{+}=t+r_{\ast},
\end{equation}
\begin{equation}  \label{3}
x^{-}=t-r_{\ast},
\end{equation}
where $r_{\ast}$ and $r$ are related by the following equation
\begin{equation}  \label{4}
\frac{dr}{dr_{\ast}}=\sqrt{-\frac{g_{00}}{g_{rr}}}=\frac{1}{a(t)}%
\sqrt{1-\frac{b(r)}{r}},
\end{equation}
and $x^{+}$ corresponds to the outgoing radiation and $x^{-}$ to the
ingoing radiation. Using Eqs. (\ref{2})-(\ref{4}), (\ref{1}) can be written as
\begin{equation}
ds^{2}=2g_{+-}dx^{+}dx^{-}+R^{2}d\Omega^{2} ,  \label{5}
\end{equation}
where $R$ and $g_{+-}=-1/2$ are functions of the null coordinates $(x^{+}, x^{-})$, that correspond to the two preferred null normal directions for the symmetric
spheres $\partial_{\pm}=\partial / \partial x^{\pm}$, and $R=a(t)r$ is the
so-called areal radius \cite{R15} and $d\Omega^{2}$ is the metric for
the unit 2-sphere. Now, we define the expansions as
\begin{equation}
\Theta_{\pm}=\frac{2}{R}\partial_{\pm}R .  \label{6}
\end{equation}
These expansions tell us whether the light rays are expanding $(\Theta >0)$ or contracting $(\Theta <0)$, or equivalently area of the sphere increases or decreases in the null directions. Since the sign of $\Theta_{+}\Theta_{-}$ is invariant, a sphere is trapped
if $\Theta_{+}\Theta_{-}>0$, which yields
\begin{equation}\label{}
  H^{2}R^{2}-1+\frac{ab}{R}>0,
\end{equation}
untrapped if $\Theta_{+}\Theta_{-}<0$, yielding
\begin{equation}\label{}
  H^{2}R^{2}-1+\frac{ab}{R}<0,
\end{equation}
or
marginal if $\Theta_{+}\Theta_{-}=0$, giving
\begin{equation}\label{}
  H^{2}R^{2}-1+\frac{ab}{R}=0
\end{equation}
where $H\equiv \dot{a} / a$ is the Hubble parameter. For fixed $\Theta_{+}>0$ and $\Theta_{-}<0$, $\partial_{+}$ is also fixed outgoing and $\partial_{-}$ ingoing null normal vector. A surface which is foliated by
marginal spheres is known as a trapping horizon. In this paper for the trapping horizon $R_{h}=a(t)r_{h}$, we choose
\begin{equation}  \label{7}
\Theta_{+}|_{h}=0,
\end{equation}
which gives
\begin{equation}  \label{8}
\dot{R}_{h}+\sqrt{1-\frac{ab}{R_{h}}}=0.
\end{equation}
Note that unlike the static Morris-Thorne wormhole, the trapping horizon and
the throat of a dynamical wormhole do not coincide. In the case of static Morris-Thorne wormhole the trapping horizon is given by $b(r_{0})=r_{0}$ which is also the value of the shape function at the throat \cite{R11}.
But in our case, because of the presence of the scaling factor $a(t)$, they do not coincide. This trapping horizon is
future if $\Theta_{-}<0$ (or equivalently $\partial _{-}R<0$), giving
\begin{equation}\label{}
  \dot{a}r< \sqrt{1-\frac{b}{r}},
\end{equation}
past if $\Theta_{-}>0$ (or equivalently $\partial _{-}R>0$), giving
\begin{equation}\label{}
  \dot{a}r> \sqrt{1-\frac{b}{r}},
\end{equation}
and bifurcating if $%
\Theta_{-}=0$ (or equivalently $\partial _{-}R=0$), giving
\begin{equation}\label{}
  \dot{a}r=\sqrt{1-\frac{b}{r}}.
\end{equation}
 Further, this trapping horizon is outer if $\partial_{-}\Theta_{+}<0$, giving
 \begin{equation}\label{}
   \frac{\dot{H}}{2}+H^{2}-\frac{(b-r\acute{b})}{4a^{2}r^{3}}<0,
 \end{equation}
inner if $\partial_{-}\Theta_{+}>0$, giving
\begin{equation}\label{}
  \frac{\dot{H}}{2}+H^{2}-\frac{(b-r\acute{b})}{4a^{2}r^{3}}>0,
\end{equation}
or degenerate if $%
\partial_{-}\Theta_{+}=0$, giving
\begin{equation}\label{}
  \frac{\dot{H}}{2}+H^{2}-\frac{(b-r\acute{b})}{4a^{2}r^{3}}=0.
\end{equation}

\section{Generalized surface gravity}

In spherically symmetric spacetimes the active gravitational energy is the
Misner-Sharp energy in spaces. It reduces to Newtonian mass in the Newtonian limit for a perfect fluid. It gives Schwarzschild energy in vacuum. At null and spatial infinity it yields Bondi-Sachs, $E_{BS}$, and Arnowitt-Deser-Misner, $E_{ADM}$, energies, respectively \cite{R14}. The Misner-Sharp energy can be expressed as \cite{R21}
\begin{equation}
E=\frac{1}{2}R(1-\partial^{a}R\partial_{a}R)=\frac{R}{2}(1-2g^{+-}%
\partial_{+}R\partial_{-}R),  \label{9}
\end{equation}
which gives
\begin{equation}  \label{10}
E=\frac{R}{2}\left[\dot{R}^{2}+\frac{ab}{R}\right].
\end{equation}
On a trapping horizon this expression reads $E=R_{h}/2$.

Now the Einstein's equations of interest in local
coordinates are
\begin{equation}  \label{11}
\partial_{\pm}\Theta_{\pm}=-\frac{1}{2}\Theta^{2}_{\pm}+\Theta_{\pm}%
\partial_{\pm}\log(-g_{+-})-8\pi T_{\pm\pm},
\end{equation}
\begin{equation}  \label{12}
\partial_{\pm}\Theta_{\mp}=-\Theta_{+}\Theta_{-}+\frac{1}{R^{2}}g_{+-}+8\pi
T_{+-}.
\end{equation}
\begin{equation}\label{12a}
  \partial_{+}\Theta_{-}+\partial_{-}\Theta_{+}+\Theta_{+}\Theta_{-}=-8\pi p_{t}
\end{equation}

In non-stationary spherically symmetric spacetimes we use Kodama vector $%
K$ instead of Killing vector which was introduced by Kodama \cite{R22} and which
reduces to a Killing vector in stationary cases. The Kodama vector in null
coordinates is given by
\begin{equation}  \label{13}
K=-g^{+-}(\partial_{+}R\partial_{-}-\partial_{-}R\partial_{+}),
\end{equation}
which for spacetime (\ref{1}) in covariant form becomes
\begin{equation}  \label{14}
K_{\pm}=-\frac{1}{2}\left(\pm \dot{R}+\sqrt{1-\frac{ab}{R}}\right).
\end{equation}
The norm of $K$ is
\begin{equation}  \label{15}
\|K\|^{2}=\frac{2E}{R}-1.
\end{equation}
Note that $\|K\|^{2}=0$ on the trapping horizon $R_{h}$.

The trapping horizon is provided by this Kodama vector which is null on a
hypersurface $\partial_{+}R=0$. In our case of dynamical spacetime, the trapping horizon and
the Kodama vector play the same roles as the Killing horizon and the Killing vector
play in the static case. In static spacetimes the hypersurface where
the Killing vector vanishes is defined as the boundary of the spacetime but here
in dynamical spacetimes we use Kodama vector instead. In the above, $E$ is the Noether
charge of Kodama vector. Kodama vector and Killing vector have some similar
properties in dynamical and static spacetimes \cite{R14}, respectively. Now,
the generalized surface gravity $\kappa$ on a trapping horizon can be expressed as \cite{R23}
\begin{equation}  \label{16}
K^{a}\nabla _{[b}K_{a]}=\pm \kappa K_{b}.
\end{equation}
For metric (\ref{1}) the surface gravity on trapping horizon becomes
\begin{equation}  \label{17}
\kappa = -\frac{\ddot{R_{h}}}{2}+\frac{1}{4R_{h}^{2}}\left(ab-b^{\prime}R_{h}\right),
\end{equation}
which on using Einstein's field equations (\ref{11}) and (\ref{12}) can be written as
\begin{equation}\label{17a}
  \kappa =-\ddot{R_{h}}%
-2\pi R_{h}\left(\rho +p_{r}\right), 
\end{equation}
and
\begin{equation}\label{17b}
  \kappa =\frac{E}{R_{h}^{2}}-4\pi R_{h}\omega.
\end{equation}

This surface gravity, from Eq. (\ref{16}), equivalently, can also be expressed
as
\begin{equation}  \label{18}
\kappa =\frac{1}{2}g^{ab}\partial_{a} \partial_{b}R,
\end{equation}
on a trapping horizon. It follows that $\kappa <0$, $\kappa=0$ and $\kappa>0$
for inner, degenerate and outer trapping horizons, respectively. As mentioned above, in dynamical spherical spacetimes the Kodama
vector is the analogue of a time-like Killing vector. We cannot define surface gravity in dynamical wormholes
using Killing vector because it does not vanish everywhere. But still we
can use Kodama vector instead and define the generalized surface gravity for
static as well as dynamical traversable wormhole at a trapping horizon.
The Hawking temperature \cite{R11, R12} is $T=-\kappa_{h}/2\pi$ which, in our case from Eq. (\ref{17}), becomes
\begin{equation}  \label{19}
T=-\frac{\kappa|_{h}}{2\pi}=-\frac{1}{2\pi}\left[-\frac{\ddot{R_{h}}}{2}%
+\frac{ab-R_{h}b^{\prime}}{%
4R_{h}^{2}}\right],
\end{equation}
which is negative for the outer trapping horizon since $\kappa |_{h}>0$. It means the particles coming out of a wormhole have the same properties as that of a phantom energy because this energy is linked with negative temperature as well. Or, we can say that the phantom energy is responsible for this negative temperature \cite{R24}.

\section{Wormholes of different shapes for a power-law cosmological model}

In this section we discuss different cases using specific values of
shape functions and a particular cosmological model. We take the scale factor $%
a(t)=a_{0}t^{n}$ where $a_{0}$ and $n$ are constants. For $n=2/3$ and $n=1/2$
this scale factor represents the matter dominated universe and the radiation
dominated universe, respectively. Using this scale factor we discuss
three cases for different expressions of the shape function.


\bigskip
\begin{center}\textbf{Shape function} $b(r)=r_{0}^{2}/r$\end{center}
\bigskip

Here, for the scale factor $a(t)=a_{0}t^{n}$, we take \cite{R20} the shape
function $b(r)=r_{0}^{2}/r$. This shape function satisfies the necessary conditions which have been discussed in the beginning to have a stable wormhole solution. Using this shape function Eq. (\ref{1aa}) becomes
\begin{equation}\label{1bb}
  Z(r)=\pm r_{0}\ln \frac{r+\sqrt{r^{2}-r_{0}^{2}}}{r_{0}} .
\end{equation}
This embedding function $Z(r)$ is depicted in Figure 1 where we have set $r_{0}=1$.
\begin{figure}[ht!]
\centering
\includegraphics[scale=0.4]{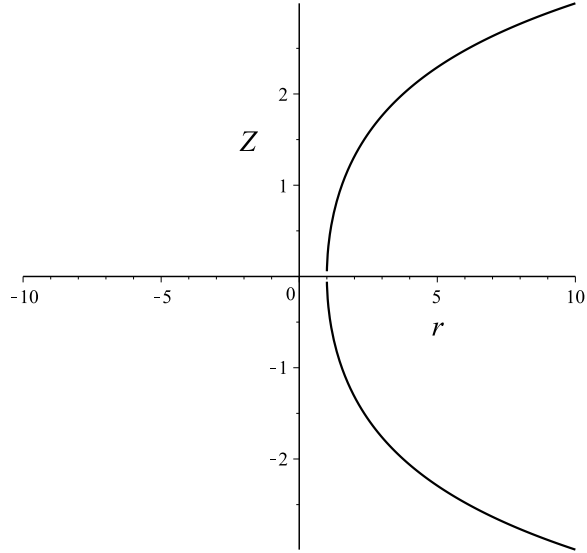}
\caption{Embedding function $Z(r)$ from Eq. (\ref{1bb}) for $r_{0}=1$.}
\label{Fig. 1}
\end{figure}
In this case, we note that, the Kodama vector from Eq. (\ref{14}) takes the form
\begin{equation}  \label{19a}
K_{\pm}=-\frac{1}{2}\left[\pm a_{0}rnt^{n-1}+\sqrt{1-\frac{r_{0}^{2}}{%
r^{2}}}\right].
\end{equation}
Using this in Eq. (\ref{16}) and evaluating on the trapping horizon gives the surface gravity
\begin{equation}  \label{19b}
\kappa = -\frac{a_{0}r_{h}n(n-1)t^{n-2}}{2}+\frac{r_{0}^{2}}{2a_{0}r_{h}^{3}t^{n}}.
\end{equation}
\begin{figure}[ht!]
\centering
\includegraphics[scale=0.4]{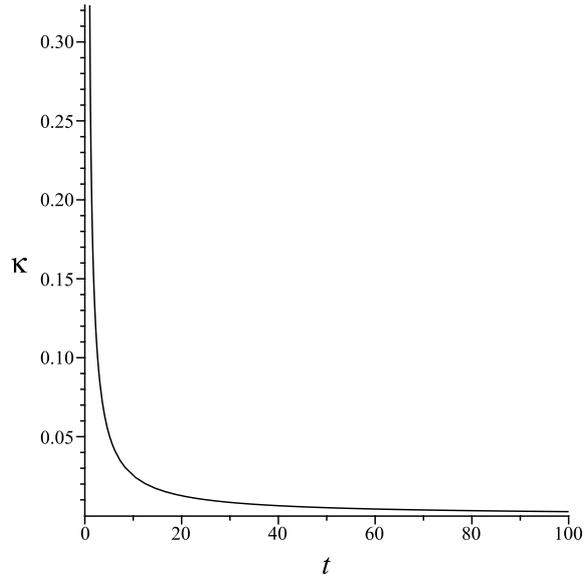}
\caption{Surface gravity as a function of time [Eq. (\ref{19b})] on the trapping horizon for $a_{0}=r_{0}=1$ and $n=1/2$}.
\label{Fig. 2}
\end{figure}

We have plotted the graph of surface gravity as a function of time in Figure 2. We see that
on the trapping horizon the value of surface gravity decreases as time
increases but never becomes equal to zero. Thus, it is positive
for all values of time.

\bigskip
\begin{center}\textbf{Shape function} $b(r)=\protect\sqrt{r_{0}r}$\end{center}
\bigskip

Here, for the scale factor $a(t)=a_{0}t^{n}$, we consider \cite{R20} the shape
function $b(r)=\sqrt{r_{0}r}$. The necessary conditions for a stable wormhole solution are satisfied by this shape function. The embedding function in this case from Eq. (\ref{1aa}) takes the form
\begin{equation}\label{2bb}
  Z(r)=\pm \frac{4 (r_{0})^{1/4}}{3}\left[(\sqrt{r}-\sqrt{r_{0}})^{3/2}+3\sqrt{r_{0}}(\sqrt{r}-\sqrt{r_{0}})^{1/2}\right].
\end{equation}
The embedding diagram for this shape function is shown in Figure 3, where we have set $r_{0}=1$.
\begin{figure}[ht!]
\centering
\includegraphics[scale=0.4]{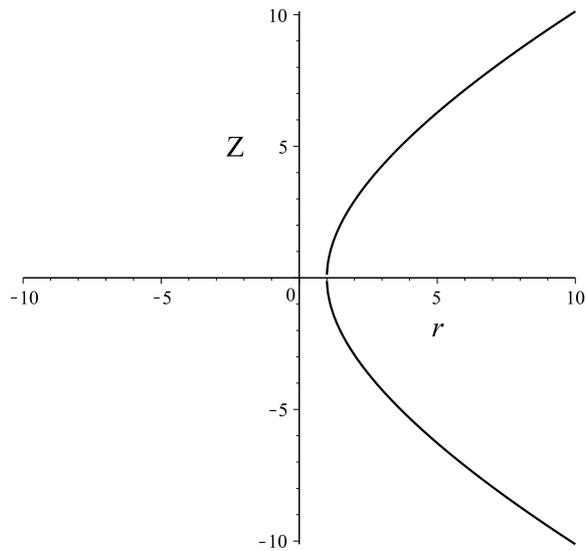}
\caption{Embedding function $Z(r)$ from Eq. (\ref{2bb}) for $r_{0}=1$.}
\label{Fig. 3}
\end{figure}

The Kodama vector, in this case, from Eq. (\ref{14}) becomes
\begin{equation}  \label{19c}
K_{\pm}=-\frac{1}{2}\left(\pm a_{0}rnt^{n-1}+\sqrt{1-\sqrt{\frac{r_{0}}{r}%
}}\right).
\end{equation}
Using this in Eq. (\ref{16}) and evaluating on the trapping horizon yields the surface gravity
\begin{equation}  \label{19d}
\kappa = -\frac{a_{0}r_{h}n\left(n-1\right)t^{n-2}}{2}+\frac{1}{8a_{0}t^{n}}\sqrt{\frac{r_{0}}{r_{h}}%
}.
\end{equation}

\bigskip
\begin{center}\textbf{Shape function} $b(r)=r_{0}(\frac{r}{r_{0}})^{\protect\gamma}$, $0\leq \protect\gamma <1$\end{center}
\bigskip

Now, we assume the scale factor $a(t)=a_{0}t^{n}$, and the shape
function $b(r)=r_{0}(\frac{r}{r_{0}})^{\gamma}$, $0\leq \gamma <1$. The embedding function in this case for $\gamma =0$ from Eq. (\ref{1aa}) is given as
\begin{equation}\label{3bb}
  Z(r)=\pm 2\sqrt{r_{0}(r-r_{0})}.
\end{equation}
The graph of this function is shown in Figure 4, where we have taken $r_{0}=1$
\begin{figure}[ht!]
\centering
\includegraphics[scale=0.4]{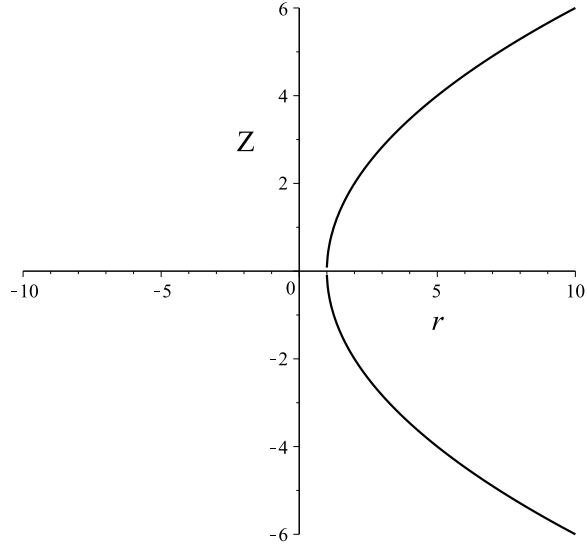}
\caption{Embedding function $Z(r)$ from Eq. (\ref{3bb}) for $r_{0}=1$}
\label{Fig. 4}
\end{figure}

The Kodama vector takes the form
\begin{equation}  \label{19e}
K_{\pm}=-\frac{1}{2}\left(\pm a_{0}rnt^{n-1}+\sqrt{1-\left(\frac{r}{r_{0}}%
\right)^{\gamma -1}}\right).
\end{equation}
In this case the surface gravity, from Eq. (\ref{16}), on the trapping horizon becomes
\begin{equation}  \label{19f}
\kappa =-\frac{a_{0}r_{h}n\left(n-1\right)t^{n-2}}{2}+\frac{1-\gamma}{4a_{0}r_{0}t^{n}}\left(%
\frac{r_{h}}{r_{0}}\right)^{\gamma -2}.
\end{equation}

\section{Wormholes of different shapes for an exponential cosmological model}

In this section we discuss wormholes with different shape functions in the framework of the cosmological model with the scale factor $a(t)=a_{0}e^{t}$ where $a_{0}$ is constant. This scale factor represents inflating wormhole \cite{R25}. The exponential scale
factor is the consequence of positive vacuum energy. In this cosmology, we discuss three cases of specific forms of
wormhole shape functions.


\bigskip
\begin{center}\textbf{Shape function} $b(r)=r_{0}^{2}/r$\end{center}
\bigskip

Here, for the scale factor $a(t)=a_{0}e^{t}$, we consider the shape
function $b(r)=r_{0}^{2}/r$. In this case the components of
the Kodama vector become
\begin{equation}  \label{19g}
K_{\pm}=-\frac{1}{2}\left[\pm a_{0}re^{t}+\sqrt{1-\frac{r_{0}^{2}}{r^{2}}} \right].
\end{equation}
Using this in Eq. (\ref{16}) and evaluating on the trapping horizon yields the surface gravity as
\begin{equation}  \label{19h}
\kappa = -\frac{a_{0}r_{h}e^{t}}{2}+\frac{r_{0}^{2}}{2a_{0}r_{h}^{3}e^{t}}.
\end{equation}


\bigskip
\begin{center}\textbf{Shape function} $b(r)=\protect\sqrt{r_{0}r}$\end{center}
\bigskip

If we assume the scale factor $a(t)=a_{0}e^{t}$, and the shape
function $b(r)=\sqrt{r_{0}r}$, the Kodama
vector becomes
\begin{equation}  \label{19i}
K_{\pm}=-\frac{1}{2}\left(\pm a_{0}re^{t}+\sqrt{1-\sqrt{\frac{r_{0}}{r}}}\right).
\end{equation}
Using Eq. (\ref{16}), the surface gravity on the trapping horizon takes the form
\begin{equation}  \label{19j}
\kappa = -\frac{a_{0}r_{h}e^{t}}{2}+\frac{1}{8a_{0}r_{h}e^{t}}\sqrt{\frac{r_{0}}{%
r_{h}}}.
\end{equation}


\bigskip
\begin{center}\textbf{Shape function} $b(r)=r_{0}(\frac{r}{r_{0}})^{\protect\gamma}$, $0<\protect\gamma <1$\end{center}
\bigskip

In this case we take the shape
function $b(r)=r_{0}(\frac{r}{r_{0}})^{\gamma}$ and the same scale factor as in the previous example. Here the Kodama vector is given by
\begin{equation}  \label{19k}
K_{\pm}=-\frac{1}{2}\left(\pm a_{0}re^{t}+\sqrt{1-\left(\frac{r}{r_{0}}%
\right)^{\gamma -1}}\right).
\end{equation}
From Eq. (\ref{16}) the surface gravity on the trapping horizon becomes
\begin{equation}  \label{19l}
\kappa =-\frac{a_{0}r_{h}e^{t}}{2}+\frac{1-\gamma}{4a_{0}r_{0}e^{t}}\left(\frac{%
r_{h}}{r_{0}}\right)^{\gamma -2}.
\end{equation}

\section{Generalized surface gravity for wormholes with and without the cosmological constant}

In this section we consider wormholes of different shapes in different cosmologies with and without the cosmological constant $\Lambda$. We will analyze these for anisotropic fluid where radial and tangential pressures satisfy $p_{r}=\omega _{r}\rho$ and $p_{t}=\omega_{t}\rho$. Clearly for $\omega_{r}=\omega_{t}$ pressure becomes isotropic.

\bigskip
\begin{center}\textbf{Static wormholes}\end{center}
\bigskip

Here we discuss static wormholes for cosmological constant ($\Lambda =0$). In the static case ($a(t)=1$) we take shape function $b(r)=r_{0}(\frac{r}{r_{0}})^{-1/\omega _{r}}$. Here $\omega _{r}$ is a constant state parameter, satisfying $p_{r}=\omega _{r}\rho$ and $p_{t}=-\frac{1}{2}(1+\omega _{r})\rho$, where $p_{r}$ and $p_{t}$ are radial and tangential pressures while $\rho$ is the energy density. For this case the wormhole metric takes the form \cite{R24A}

\begin{equation}\label{}
  ds^{2}=-dt^{2}+\frac{dr^{2}}{1-(r/r_{0})^{-(1+\omega _{r})/\omega _{r}}}+r^{2}d\Omega ^{2}.
\end{equation}
In the range $\omega _{r}<-1$, we have asymptotically flat wormhole metric with positive energy density while for $\omega _{r}>0$ the energy density becomes negative but still we have an asymptotically flat wormhole. This static traversable wormhole was first considered in Ref. \cite{R10}.
In the static case we have a bifurcating trapping horizon on the wormhole throat location, given by Eq. (\ref{7}) as
\begin{equation}\label{}
  r_{h}=r_{0}.
\end{equation}
The Kodama vector in this case becomes
\begin{equation}\label{}
  K_{\pm}=-\frac{1}{2}\sqrt{1-(r/r_{0})^{-(1+\omega _{r})/\omega _{r}}}.
\end{equation}
Finally, the surface gravity from Eq. (\ref{16}) when evaluated on the trapping horizon $r=r_{h}=r_{0}$ takes the form
\begin{equation}\label{}
  \kappa =\frac{1+\omega _{r}}{4r_{0}\omega _{r}}.
\end{equation}

\bigskip
\begin{center}\textbf{Evolving wormholes with $\Lambda =0$}\end{center}
\bigskip

We discuss a non-static wormhole  with shape function
\begin{equation}\label{S5}
  b(r)=r_{0}(\frac{r}{r_{0}})^{-1/\omega _{r}}+kr_{0}^{3}(\frac{r}{r_{0}})^{3}-kr_{0}^{3}(\frac{r}{r_{0}})^{-1/\omega _{r}},
\end{equation}
in the background of a cosmology with the scale factor $a(t)= t \sqrt{-k}+F$, where $k$ and $F$ are constants and $\omega _{r}$ satisfies the same conditions as discussed above for the static case. This shape function also satisfies the near throat conditions discussed earlier. With these values the wormhole metric can be written as \cite{R24A}
\begin{equation}\label{}
  ds^{2}=-dt^{2}+(\sqrt{-k}t+F)^{2}\left(\frac{dr^{2}}{1-(r/r_{0})^{-(1+\omega _{r})/\omega _{r}}-kr_{0}^{2}(\frac{r}{r_{0}})^{2}+kr_{0}^{2}(\frac{r}{r_{0}})^{-(1+\omega _{r})/\omega _{r}}}+r^{2}d\Omega ^{2}\right).
\end{equation}
Here $k=-1, 0, +1$ correspond to open, flat and closed universe, respectively. In the above case must have $k\leq 0$ for preserving the Lorentzian signatures. Otherwise, for $k>0$ the signatures changes to the Euclidean one giving rise to Euclidean wormholes. The trapping horizon for this metric is given by the expression
\begin{equation}\label{}
  \sqrt{-k}+\sqrt{1-(r_{h}/r_{0})^{-(1+\omega _{r})/\omega _{r}}-kr_{0}^{2}(\frac{r_{h}}{r_{0}})^{2}+kr_{0}^{2}(\frac{r_{h}}{r_{0}})^{-(1+\omega _{r})/\omega _{r}}}=0,
\end{equation}
whereas the Kodama vector in the component form becomes
\begin{equation}\label{}
  K_{\pm}=-\frac{1}{2}\left(\pm \sqrt{-k}r+\sqrt{1-(r/r_{0})^{-(1+\omega _{r})/\omega _{r}}-kr_{0}^{2}(\frac{r}{r_{0}})^{2}+kr_{0}^{2}(\frac{r}{r_{0}})^{-(1+\omega _{r})/\omega _{r}}}\right).
\end{equation}
Finally, the surface gravity from Eq. (\ref{16}) on trapping horizon takes the form
\begin{equation}\label{}
  \kappa =\frac{1}{4r_{h}^{2}(\sqrt{-k}t+F)}\left[\frac{(1+\omega _{r})r_{0}(1-kr_{0}^{2})}{\omega _{r}}(\frac{r_{h}}{r_{0}})^{-1/\omega _{r}}-2kr_{0}^{3}(\frac{r_{h}}{r_{0}})^{3}\right].
\end{equation}


\bigskip
\begin{center}\textbf{Inflating de Sitter wormholes}\end{center}
\bigskip

When we include the cosmological constant, the wormholes do not remain asymptotically flat and the expansion of the wormhole is accelerated. Here we discuss  a case of exponential scale factor $a(t)=a_{0}e^{\pm \sqrt{\Lambda /3}t}$ for $\Lambda > 0$. For this scale factor we take the shape function $b(r)=r_{0}(\frac{r}{r_{0}})^{-1/\omega _{r}}$, so that the wormhole metric takes the form
\begin{equation}\label{}
  ds^{2}=-dt^{2}+a_{0}^{2}e^{\pm 2\sqrt{\Lambda /3}t}\left[\frac{dr^{2}}{1-(r/r_{0})^{-(1+\omega _{r})/\omega _{r}}}+r^{2}d\Omega ^{2}\right],
\end{equation}
describing contracting and expanding wormholes. The positive sign in this scale factor represents inflation giving exponential expansion of an inflating wormhole. These wormholes were first considered in Ref.  \cite{R25}. This wormhole is asymptotically de Sitter for $\omega _{r}<-1$ with positive energy density everywhere, while for $\omega _{r}>0$ the energy density is negative everywhere and the wormhole solution is still asymptotically de Sitter universe. When $\Lambda $ vanishes we obtain the static case discussed earlier. For these wormholes the trapping horizon is given by the expression
\begin{equation}\label{}
  \pm a_{0}\sqrt{\Lambda /3}e^{\pm \sqrt{\Lambda /3}t}r_{h}+\sqrt{1-(r_{h}/r_{0})^{-(1+\omega _{r})/\omega _{r}}}=0,
\end{equation}
whereas the Kodama vector in the component form is given by
\begin{equation}\label{}
  K_{\pm}=-\frac{1}{2}\left(\pm a_{0}(\pm \sqrt{\Lambda /3})e^{\pm \sqrt{\Lambda /3}t}r+\sqrt{1-(r/r_{0})^{-(1+\omega _{r})/\omega _{r}}}\right),
\end{equation}
yielding the surface gravity
\begin{equation}\label{}
  \kappa =-\frac{a_{0}r_{h}\Lambda e^{\pm \sqrt{\Lambda /3}t}}{6}+\frac{r_{0}(1+\omega _{r})}{4a_{0}\omega _{r}r_{h}^{2}e^{\pm \sqrt{\Lambda /3}t}}(\frac{r_{h}}{r_{0}})^{-1/\omega _{r}}.
\end{equation}

\bigskip
\begin{center}\textbf{Evolving de Sitter wormholes in closed universe}\end{center}
\bigskip

Now we discuss the more general case when $\Lambda \neq 0$, and the shape function is given by Eq. (\ref{S5}). As the cosmological constant is nonzero, the wormhole is not asymptotically flat. For different values of constant $k$ we can have different kinds of scale factors discussed in detail in Ref. \cite{R18}. For $k=1$ and $\Lambda > 0$, we take the scale factor given by $a(t)=\sqrt{\frac{3}{\Lambda}}\cosh(\sqrt{\frac{\Lambda}{3}}t+\phi _{0})$ where $\phi _{0}$ is a constant. With these values the de Sitter wormhole of a closed universe becomes
\begin{equation}\label{}
  ds^{2}=-dt^{2}+\frac{3}{\Lambda}\cosh^{2}(\sqrt{\frac{\Lambda}{3}}t+\phi _{0})\left(\frac{dr^{2}}{1-(r/r_{0})^{-(1+\omega _{r})/\omega _{r}}-r_{0}^{2}(\frac{r}{r_{0}})^{2}+r_{0}^{2}(\frac{r}{r_{0}})^{-(1+\omega _{r})/\omega _{r}}}+r^{2}d\Omega ^{2}\right).
\end{equation}
The trapping horizon for this wormhole is given by the expression
\begin{equation}\label{}
  \sinh(\sqrt{\frac{\Lambda}{3}}t+\phi _{0})r_{h}+\sqrt{1-(r_{h}/r_{0})^{-(1+\omega _{r})/\omega _{r}}-r_{0}^{2}(\frac{r_{h}}{r_{0}})^{2}+r_{0}^{2}(\frac{r_{h}}{r_{0}})^{-(1+\omega _{r})/\omega _{r}}}=0,
\end{equation}
and the Kodama vector in the component form is given by
\begin{equation}\label{}
  K_{\pm}=-\frac{1}{2}\left(\pm \sinh(\sqrt{\frac{\Lambda}{3}}t+\phi _{0})r+\sqrt{1-(r/r_{0})^{-(1+\omega _{r})/\omega _{r}}-r_{0}^{2}(\frac{r}{r_{0}})^{2}+r_{0}^{2}(\frac{r}{r_{0}})^{-(1+\omega _{r})/\omega _{r}}}\right).
\end{equation}
Evaluating Eq. (\ref{16}) on the trapping horizon gives for the surface gravity
\begin{eqnarray} 
  \kappa &=& -\frac{\sqrt{\Lambda}r_{h}}{2\sqrt{3}}\cosh(\sqrt{\frac{\Lambda}{3}}t+\phi _{0}) \nonumber \\ &-&\frac{\sqrt{\Lambda}}{4r_{h}^{2}\sqrt{3}\cosh(\sqrt{\frac{\Lambda}{3}}t+\phi _{0})}\left[\frac{(1+\omega _{r})r_{0}(1-r_{0}^{2})}{\omega _{r}}(\frac{r_{h}}{r_{0}})^{-1/\omega _{r}}-2r_{0}^{3}(\frac{r_{h}}{r_{0}})^{3}\right].
\end{eqnarray}

\bigskip
\begin{center}\textbf{Evolving de Sitter wormholes in open universe}\end{center}
\bigskip

If in Eq. (\ref{S5}) we take $k=-1$ then for $\Lambda > 0$ the scale factor is given by $a(t)=\sqrt{\frac{3}{\Lambda}}\sinh(\sqrt{\frac{\Lambda}{3}}t+\phi _{0})$ and the wormhole metric takes the form
\begin{equation}\label{}
  ds^{2}=-dt^{2}+\frac{3}{\Lambda}\sinh^{2}(\sqrt{\frac{\Lambda}{3}}t+\phi _{0})\left(\frac{dr^{2}}{1-(r/r_{0})^{-(1+\omega _{r})/\omega _{r}}+r_{0}^{2}(\frac{r}{r_{0}})^{2}-r_{0}^{2}(\frac{r}{r_{0}})^{-(1+\omega _{r})/\omega _{r}}}+r^{2}d\Omega ^{2}\right).
\end{equation}
In this case the expression for the trapping horizon is
\begin{equation}\label{}
  \cosh(\sqrt{\frac{\Lambda}{3}}t+\phi _{0})r_{h}+\sqrt{1-(r_{h}/r_{0})^{-(1+\omega _{r})/\omega _{r}}+r_{0}^{2}(\frac{r_{h}}{r_{0}})^{2}-r_{0}^{2}(\frac{r_{h}}{r_{0}})^{-(1+\omega _{r})/\omega _{r}}}=0
\end{equation}
and the Kodama vector takes the form
\begin{equation}\label{}
  K_{\pm}=-\frac{1}{2}\left(\pm \cosh(\sqrt{\frac{\Lambda}{3}}t+\phi _{0})r+\sqrt{1-(r/r_{0})^{-(1+\omega _{r})/\omega _{r}}+r_{0}^{2}(\frac{r}{r_{0}})^{2}-r_{0}^{2}(\frac{r}{r_{0}})^{-(1+\omega _{r})/\omega _{r}}}\right).
\end{equation}
Thus surface gravity on trapping horizon becomes
\begin{eqnarray}\label{}
  \kappa &=&-\frac{\sqrt{\Lambda}r_{h}}{2\sqrt{3}}\sinh(\sqrt{\frac{\Lambda}{3}}t+\phi _{0}) \nonumber \\ &-&\frac{\sqrt{\Lambda}}{4r_{h}^{2}\sqrt{3}\sinh(\sqrt{\frac{\Lambda}{3}}t+\phi _{0})}\left[\frac{(1+\omega _{r})r_{0}(1+r_{0}^{2})}{\omega _{r}}(\frac{r_{h}}{r_{0}})^{-1/\omega _{r}}+2r_{0}^{3}(\frac{r_{h}}{r_{0}})^{3}\right].
\end{eqnarray}

\bigskip
\begin{center}\textbf{Evolving anti-de Sitter wormholes in open universe}\end{center}
\bigskip

Finally we discuss a case of negative cosmological constant ($\Lambda < 0$) with $k=-1$ in Eq. (\ref{S5}). We take the scale factor as $a(t)=\sqrt{\frac{-3}{\Lambda}}sin(\sqrt{\frac{-\Lambda}{3}}t+\phi _{0})$, so that the wormhole metric can be written as
\begin{eqnarray}\label{}
  ds^{2}&=&-dt^{2} +\frac{-3}{\Lambda}\sin^{2}(\sqrt{\frac{-\Lambda}{3}}t+\phi _{0}) \nonumber \\ &\times& \left(\frac{dr^{2}}{1-(r/r_{0})^{-(1+\omega _{r})/\omega _{r}}+r_{0}^{2}(\frac{r}{r_{0}})^{2}-r_{0}^{2}(\frac{r}{r_{0}})^{-(1+\omega _{r})/\omega _{r}}}+r^{2}d\Omega ^{2}\right).
\end{eqnarray}
Its trapping horizon is given by the expression
\begin{equation}\label{}
  \cos(\sqrt{\frac{-\Lambda}{3}}t+\phi _{0})r_{h}+\sqrt{1-(r_{h}/r_{0})^{-(1+\omega _{r})/\omega _{r}}+r_{0}^{2}(\frac{r_{h}}{r_{0}})^{2}-r_{0}^{2}(\frac{r_{h}}{r_{0}})^{-(1+\omega _{r})/\omega _{r}}}=0,
\end{equation}
and the Kodama vector takes the form
\begin{equation}\label{}
  K_{\pm}=-\frac{1}{2}\left(\pm \cos(\sqrt{\frac{-\Lambda}{3}}t+\phi _{0})r+\sqrt{1-(r/r_{0})^{-(1+\omega _{r})/\omega _{r}}+r_{0}^{2}(\frac{r}{r_{0}})^{2}-r_{0}^{2}(\frac{r}{r_{0}})^{-(1+\omega _{r})/\omega _{r}}}\right).
\end{equation}
Using all these expressions the surface gravity becomes
\begin{eqnarray}\label{}
  \kappa &=&\frac{\sqrt{-\Lambda}r_{h}}{2\sqrt{3}}\sin(\sqrt{\frac{-\Lambda}{3}}t+\phi _{0}) \nonumber \\ &-&\frac{\sqrt{-\Lambda}}{4r_{h}^{2}\sqrt{3}\sin(\sqrt{\frac{-\Lambda}{3}}t+\phi _{0})}\left[\frac{(1+\omega _{r})r_{0}(1+r_{0}^{2})}{\omega _{r}}(\frac{r_{h}}{r_{0}})^{-1/\omega _{r}}+2r_{0}^{3}(\frac{r_{h}}{r_{0}})^{3}\right].
\end{eqnarray}

\section{Unified first law for dynamical wormholes}

We know that we can formulate a unified first law of thermodynamics in spherically symmetric spacetimes \cite{R15}. This law describes the gradient of the active gravitational energy, using Einstein's field equations, as a sum of two terms, the energy supply term and the work term. When we project this along the trapping horizon we get the first law of wormhole dynamics. This expression involves the area and surface gravity and has the same form as the wormhole statics if we replace the perturbations by the derivative along the trapping horizon. For the first law of wormhole dynamics we need to define the generalized surface gravity using Kodama vector and trapping horizon in the same manner as the first law of wormhole statics requires the stationary definition of surface gravity using Killing vector and Killing horizon.  Also, this expression involves energy at horizon rather than at infinity.

Using the energy-momentum tensor of the background fluid we construct a function and a vector in the local coordinates as
\begin{equation}
\omega =-g_{+-}T^{+-}=\frac{\rho -p_{r}}{2},  \label{20}
\end{equation}
and
\begin{equation}
\psi =T^{++}\partial_{+}R\partial_{+}+T^{--}\partial_{-}R\partial_{-} .
\label{21}
\end{equation}
In components form it can be written as
\begin{equation}  \label{21}
\psi _{\pm}=\left(\frac{\rho +p_{r}}{4}\right)\left(-\dot{R}\pm \sqrt{1-\frac{ab}{R}}\right).
\end{equation}
Now the unified first law of thermodynamics can be written by taking gradient of the gravitational energy and using Einstein's field equations as \cite{R15}
\begin{equation}  \label{22}
\partial_{\pm}E=A\psi_{\pm}+\omega \partial_{\pm}V,
\end{equation}
with
\begin{equation}  \label{23}
\partial_{\pm}E=2\pi R^{2}\left(\pm\rho \sqrt{1-\frac{ab}{R}}-\dot{R}p_{r}\right),
\end{equation}
where $A=4\pi R^{2}$ and $V=4\pi R^{3}/3$ are the area and areal
volume of the spheres of symmetry and the corresponding flat space,
respectively. We can interpret $\omega$ and $\psi$ physically as the energy density and the energy flux (outward flux minus the inward flux). The right hand side of the unified first law (\ref{22}) is the sum of two terms, the first term $A\psi_{\pm}$, called the energy supply term, produces variation in
energy of the spacetime and the second term, $\omega \partial_{\pm}V$, called the
work term, supports the spacetime structure. Finally, Eq. (\ref{22})
when projected along the trapping horizon gives the first law of wormhole dynamics which can be expressed as
\begin{equation}  \label{24}
E^{\prime}=\frac{\kappa A^{\prime}}{8\pi}+\omega V^{\prime},
\end{equation}
where we have used the notation $F^{\prime}=z.\nabla F$. Here $%
z=z^{+}\partial_{+}+z^{-}\partial_{-}$ is a tangent vector to the trapping
horizon. This expression defines a relation between surface area and
geometric entropy as
\begin{equation}  \label{25}
S \propto A|_{h}.
\end{equation}
Using Eq. (\ref{19}), Eq. (\ref{24}) takes the form
\begin{equation}  \label{26}
E^{\prime}=-TS^{\prime}+\omega V^{\prime},
\end{equation}
on the trapping horizon, where
\begin{equation}  \label{27}
S=\frac{A|_{h}}{4}.
\end{equation}
The negative sign in front of the first term of the right hand side in Eq. (\ref{26})
is due to the energy removal from the wormhole. Thus the first law of wormhole dynamics is stated as:
the change in the gravitational energy is equal to the energy that is
removed from the wormhole plus the work term which is carried out in the
wormhole.

\section{Thermodynamic stability}

In this section we study the thermodynamic stability of wormholes under consideration using the variables $E, T, S, P$ and $V$. We follow the usual criterion for thermodynamic stability, that is  $\frac{\partial \bar{P}}{\partial V}\mid _{T}\leq 0$ and $C_{P}\geq C_{V}\geq 0$ \cite{R25A, R25B}, where $\bar{P}=(P_{r}+2P_{t})/3$ is the average pressure and $C_{P}$ and $C_{V}$ are specific heats at constant pressure and volume, respectively.

We subtract Eq. (\ref{17a}) from (\ref{17b}) and rearrange the terms to obtain
\begin{equation}\label{ts1}
  p_{r}=-\frac{1}{8\pi R^{2}}-\frac{\ddot{a}}{4\pi a}. 
\end{equation}
Eq. (\ref{12a}) on the trapping horizon yields
\begin{equation}\label{ts2}
  2p_{t}=\frac{\kappa}{2\pi R}. 
\end{equation}
From Eqs. (\ref{ts1}) and (\ref{ts2}), using the definition of Hawking temperature, we obtain the average pressure $\bar{P}$ as
\begin{equation}\label{ts3}
  \bar{P}=\frac{p_{r}+2p_{t}}{3}=-\frac{1}{24\pi R^{2}}-\frac{\ddot{a}}{12\pi a}-\frac{T}{3R}, 
\end{equation}
which is the equation of state in three state parameters $T,\bar{P}$ and $V$. From this equation we can analyze the thermodynamic stability of wormhole.

Stable equilibrium of a thermodynamic system requires that $\frac{\partial \bar{P}}{\partial V}\mid _{T}\leq 0$  where

\begin{equation}\label{ts4}
  \frac{\partial \bar{P}}{\partial V}\mid _{T}=\frac{(4\pi/3)^{2/3}}{36\pi V^{5/3}}+\frac{(4\pi/3)^{1/3}T}{9V^{4/3}}.
\end{equation}
Now to ensure the stable equilibrium we must have
\begin{equation}\label{ts5}
  T\leq -\frac{1}{4\pi R},
\end{equation}
thus temperature assumes negative values everywhere for stable equilibrium which is attributed to the exotic matter. From Eq. (\ref{ts3}) we have
\begin{equation}\label{ts6}
  \bar{P}\geq \frac{1}{24\pi R^{2}}-\frac{\ddot{a}}{12\pi a}.
\end{equation}
If the scale factor is a linear function of time then $\ddot{a}=0$ and then $\bar{P}$ will assume the positive values everywhere, otherwise it could be negative somewhere.

Another condition for stable equilibrium is $C_{P}\geq C_{V}\geq 0$. Now since, the constant $V$ means constant $E$ and $S$ so by the definition of $C_{V}$, 
\begin{equation}\label{ts7}
  C_{V}=\frac{\partial E}{\partial T}\mid _{V}=T\frac{\partial S}{\partial T}\mid _{V}=0,
\end{equation}
which means we can define heat capacity only at constant pressure as
\begin{equation}\label{ts8}
  C_{P}=T\frac{\partial S}{\partial T}\mid _{P}=\frac{(24a\pi \bar{P}R^{2}+2\ddot{a}R^{2}+a)2\pi R^{2}}{24a\pi \bar{P}R^{2}+2\ddot{a}R^{2}-a},
\end{equation}
where from Eq. (\ref{ts3}),
\begin{equation}\label{ts9}
  T=-(3R\bar{P}+\frac{1}{8\pi R}+\frac{\ddot{a}R}{4\pi a}).
\end{equation}
Now from Eq. (\ref{ts6}), to ensure the stable equilibrium, we can take the value of $\bar{P}$, for any non-negative $\epsilon$, as
\begin{equation}\label{ts10}
  \bar{P}= \frac{1}{24\pi R^{2}}-\frac{\ddot{a}}{12\pi a}+\epsilon. 
\end{equation}
Thus Eq. (\ref{ts8}) on using Eq. (\ref{ts10}) takes the form
\begin{equation}\label{ts11}
   C_{P}=\frac{1}{6\epsilon}+2\pi R^{2}.
\end{equation}
which is always positive. Thus the wormholes are thermodynamically stable. This means that for stable equilibrium the average pressure is always positive for linear scale factor, however it may also have negative values for non-linear scale factor while temperature is always negative as is also depicted in Ref. \cite{R25C} in which the possibility of negative temperature emerging from the exotic matter distribution was proposed.

\section{Areal radius coordinates}

Sometimes it is useful to employ areal radius $R\equiv a(t)r$ as a coordinate instead of $r$. The Schwarzschild-like coordinates are one of this kind of coordinate systems. Also, these systems provide what are called the pseudo-Painleve-Gullstrand coordinates \cite{R26}.  Using the areal radius, metric (\ref{1}) can be written in the pseudo-Painleve-Gullstrand form as
\begin{equation}  \label{28a}
ds^{2}=-\left[\frac{1-\frac{ab}{R}-R^{2}H^{2}}{1-\frac{ab}{R}}\right]dt^{2}+%
\frac{dR^{2}}{\left(1-\frac{ab}{R}\right)}-\frac{2HR}{\left(1-\frac{ab}{R}\right)}dtdR+R^{2}d\Omega^{2},
\end{equation}
where $H\equiv \dot{a}/a$ is the Hubble parameter. As required in the Painleve-Gullstrand coordinates the coefficient of $dR^{2}$ is not unity \cite{R27}.

To obtain the Schwarzschild-like form we define a new time $T$ by using
the transformation
\begin{equation}  \label{29b}
dT=\frac{1}{F}\left(dt+\beta dR\right),
\end{equation}
where $F$ is the integration factor which satisfies
\begin{equation}  \label{30c}
\frac{\partial}{\partial R}(\frac{1}{F})=\frac{\partial}{\partial t}(\frac{%
\beta}{F}).
\end{equation}
Here $\beta (t, R)$ will be chosen later. Using Eq. (\ref{29b}) in Eq. (\ref%
{28a}) implies
\begin{align}  \label{31d}
ds^{2}=&-\left[\frac{1-\frac{ab}{R}-R^{2}H^{2}}{1-\frac{ab}{R}}%
\right]F^{2}dT^{2}+\left[\frac{1+2HR\beta -\left(1-\frac{ab}{R}%
-R^{2}H^{2}\right)\beta^{2}}{1-\frac{ab}{R}}\right]dR^{2}  \notag \\
&+\left[\frac{2F\beta \left(1-\frac{ab}{R}-R^{2}H^{2}\right)-2HRF}{1-\frac{ab}{%
R}}\right]dTdR+R^{2}d\Omega^{2}.
\end{align}
The cross term $dTdR$ is eliminated if we choose
\begin{equation}  \label{32e}
\beta =\frac{HR}{1-\frac{ab}{R}-R^{2}H^{2}}.
\end{equation}
Thus metric (\ref{31d}) takes the diagonal form
\begin{equation}  \label{33f}
ds^{2}=-\left[\frac{1-\frac{ab}{R}-R^{2}H^{2}}{1-\frac{ab}{R}}%
\right]F^{2}dT^{2}+\left[\frac{1}{1-\frac{ab}{R}-R^{2}H^{2}}\right]%
dR^{2}+R^{2}d\Omega^{2},
\end{equation}
where $F=F(T,R)$, $a$ and $H$ depend on $T$ implicitly.

This metric (\ref{33f}) can be put in the form of (\ref{5}) by using
null coordinates $x^{+}=T+R_{\ast}$ and $x^{-}=T-R_{\ast}$ where
\begin{equation}
dR/dR_{\ast}=\sqrt{-\frac{g_{TT}}{g_{RR}}}=\frac{\left[1-\frac{ab}{R}%
-R^{2}H^{2}\right]F}{\sqrt{1-\frac{ab}{R}}}.
\end{equation}
The trapping horizon in this case is given by $\Theta_{+}=\frac{2}{R}%
\partial _{+}R=0$ which gives
\begin{equation}
\left(1-\frac{ab}{R}\right)=H^{2}R^{2}.
\end{equation}
Here we have bifurcating trapping horizon as $\Theta_{+}=0$ implies $\Theta_{-}=0$.

The Misner-Sharp energy, energy flux and energy density are given, respectively, by
\begin{equation}  \label{34g}
E=\frac{R}{2}\left[1-\left[1-\frac{ab}{R}-R^{2}H^{2}\right]F\right],
\end{equation}
\begin{equation}  \label{35h}
\psi _{\pm}=\pm(\rho +p_{r})\frac{\left[1-\frac{ab}{R}-R^{2}H^{2}\right]F}{%
4\sqrt{1-\frac{ab}{R}}},
\end{equation}
\begin{equation}  \label{36i}
\omega =\frac{\rho-p_{r}}{2}.
\end{equation}
It may be noted that $E=R/2$ at the trapping horizon only. Now, with the quantity
\begin{equation}  \label{37j}
\partial _{\pm}E=\pm \frac{2\pi R^{2}\rho \left[1-\frac{ab}{R}%
-R^{2}H^{2}\right]F}{\sqrt{1-\frac{ab}{R}}},
\end{equation}
the first law of thermodynamics is satisfied. The Kodama vector in this case takes the form
\begin{equation}  \label{38}
K^{\pm}=\frac{\sqrt{1-\frac{ab}{R}}}{F},
\end{equation}
with $\|K\|^{2}=0$ on the trapping horizon. The generalized surface gravity from Eq. (\ref{16}) becomes
\begin{equation}  \label{39}
\kappa |_{h}=-\frac{ab^{\prime}}{2R_{h}%
}+\frac{ab}{2R_{h}^{2}}-H^{2}R_{h},
\end{equation}
which on using Einstein's field equations takes the form
\begin{equation}  \label{}
\kappa |_{h}=-2\pi R_{h}(\rho+p_{r})=\frac{E}{R_{h}^{2}}-4\pi R_{h}\omega .
\end{equation}

\section{Conclusion}
 
In this paper we have investigated dynamical traversable wormholes, which are the time
generalization of Morris-Thorne wormholes, and studied their thermodynamics and the laws of mechanics. In dynamical spacetimes the Kodama vector and the trapping horizon replace the role of the Killing vector and
Killing horizon, respectively. The Kodama vector reduces to the Killing
vector for static vacuum case. However, this is not possible for non-vacuum
cases. There is no Killing horizon (even though we do have the Killing vector) present to find the surface gravity in
wormholes. So, we find the generalized surface gravity with the help of the trapping horizon. Our results generalize the results available in the literature for the Morris-Thorne wormholes.

We have discussed wormholes in different cosmological models, with and without the cosmological constant, for their thermodynamic properties. These include de Sitter and anti-de Sitter wormholes in open, closed and flat universes. Further, we have discussed cases of asymptotically flat and asymptotically de Sitter wormholes as well.

The unified first law of
wormhole thermodynamics is derived which is stated as `the change in the
gravitational energy equals the energy removed from the wormhole plus the
work term'. We have derived the generalized surface gravity for a dynamical
traversable wormhole at the trapping horizon. This surface gravity
is positive, negative or zero for outer, inner or degenerate trapping
horizons, respectively. When we compare the results for black holes and wormholes we get useful information about these dynamical wormholes and hence about the
exotic matter which supports the construction of these spacetimes. The
gravitational energy and the work term which are responsible for the stable
structure of spacetime appearing in the first law have same sign while the
energy supply term is negative. This means that matter content takes energy
from the spacetime and then from this energy it does work for maintaining
the structure of the wormhole unlike the situation in a black hole where the sign of the
energy supply term is positive such that it gives energy to the black hole
spacetime.

We have discussed the thermodynamic stability of wormholes and have shown that, for linear scale factor, average pressure assumes positive values everywhere ($\bar{P}\geq 1/24\pi R^{2}$) which is the natural requirement in the usual thermodynamic systems. In the case of non-linear scale factor pressure could also have negative values depending on the value of the second derivative of the scale factor which is also possible in gravitational system such as in the case of dark energy. The temperature is always negative ($T\leq -1/4\pi R$) for stable thermodynamic equilibrium which could be attributed to the exotic matter.

\section*{Acknowledgements}

A research grant from the Higher Education Commission of Pakistan under
its Project No. 6151 is gratefully acknowledged.


\begin{thebibliography}{99}
\bibitem{R1} L. Flamm, Phys. Z \textbf{17} (1916) 448 .

\bibitem{R2} H. Weyl, Philosophie der Mathematik und Naturwissenschaft,
Handbuch der Philosophie, Leibniz Verlag, Munich (1928).

\bibitem{R3} A. Einstein and N. Rosen, Phys. Rev. \textbf{48} (1935) 73.

\bibitem{R4} C. W. Misner and J. A. Wheeler, Annals Phys. \textbf{2} (1957) 525.

\bibitem{R4A} S. Coleman, Nucl. Phys. \textbf{307} (1988) 867.

\bibitem{R4B} S. B. Giddings and A. Strominger, Nucl. Phys. B \textbf{321} (1988) 481.

\bibitem{R5} M. S. Morris and K. S. Thorne, Am. J. Phys. \textbf{56} (1988) 395.

\bibitem{R6} M. S. Morris, K. S. Thorne and U. Yurtsever, Phys. Rev. Lett. \textbf{61} (1988) 1446.

\bibitem{R6A} E. J. Copeland, M. Sami and S. Tsujikawa, Int. J. Mod. Phys. D \textbf{15} (2006) 1753.

\bibitem{R7} D. J. Mortlock and R. L. Webster, Mon. Not. RoyAstron. Soc.
\textbf{319} (2000) 872; A. G. Riess et al. [Supernova Search Team Collaboration],
Astron. J. \textbf{116} (1998) 1009; S. Perlmutter et al. [Supernova Cosmology
Project Collaboration], Astrophys. J. \textbf{517} (1999) 565; J. L. Tonry et al.
[Supernova Serach Team Collaboration], Astrophys. J. \textbf{594} (2003) 1; D. N.
Spergel et al. [WMAP Collaboration], Astrophys. J. Suppl. \textbf{148} (2003) 175;
C. L. Bennett et al. Astrophys. J. Suppl. \textbf{148} (2003) 1; M. Tegmark et al.
[SDSS Collaboration], Phys. Rev. D \textbf{69} (2004) 103501.

\bibitem{R8} R. R. Cardwell, Phys. Lett. B \textbf{545}
(2002) 23; V. K. Onemli and R. P. Woodard, Class. Quantum Gravit. \textbf{19} (2002) 4607;
S. M. Carroll, M. Hoffman and M. Tredden, Phys. Rev. D \textbf{68} (2003) 023509; P.
F. Gonzelez-Diaz, Phys. Lett. B \textbf{586} (2004) 1.

\bibitem{R9} S. V. Sushkov, Phys. Rev. D \textbf{71} (2005) 043520.

\bibitem{R10} F. S. N. Lobo, Phys. Rev. D \textbf{71} (2005) 084011.

\bibitem{R11} P. Martin-Moruno and P. F. Gonzalez-Diaz, Phys. Rev. D \textbf{80} (2009) 024007.

\bibitem{R12} P. Martin-Moruno and P. F. Gonzalez-Diaz, Class. Quantum Gravit. \textbf{26} (2009) 215010.

\bibitem{R13} S. A. Hayward, Phys. Rev. D \textbf{49} (1994) 6467.

\bibitem{R14} S. A. Hayward, Phys. Rev. D \textbf{53} (1996) 1938.

\bibitem{R15} S. A. Hayward, Class. Quantum Gravit. \textbf{15} (1998) 3147.

\bibitem{R16} S. A. Hayward, Phys. Rev. D \textbf{70} (2004) 104027.

\bibitem{R23} S. W. Hawking, R. D. Criscienzo, M. Nadalini, L. Vanzo and S. Zerbini, Class. Quantum Gravit. \textbf{26} (2009) 062001.

\bibitem{R17} S. A. Hayward, Int. J. Mod. Phys. D \textbf{8} (1999) 373.

\bibitem{R18} M. Cataldo, S. del Campo, P. Minning and P. Salgado, Phys. Rev. D \textbf{79} (2009) 024005.

\bibitem{R25} T. A. Roman, Phys. Rev. D \textbf{47} (1993) 1370.

\bibitem{R19} N. M. Garcia and F. S. N. Lobo, Phys. Rev. D \textbf{82} (2010) 104018.

\bibitem{R20} F. S. N. Lobo and M. A. Oliveira, Phys. Rev. D \textbf{80} (2009) 104012.

\bibitem{LL} J. P. S. Lemos and F. S. N. Lobo, Phys. Rev. D \textbf{78} (2008) 044030.

\bibitem{GGS} J. A. Gonzalez, F. S. Guzman and O. Sarbach, Phys. Rev. D \textbf{80} (2009) 024023.

\bibitem{GGS1} J. A. Gonzalez, F. S. Guzman and O. Sarbach, Class. Quantum Gravit. \textbf{26} (2009) 015010.

\bibitem{RT} R. Troncoso, J. High Energy Phys.  \textbf{8} (2008) 081.


\bibitem{R21} C. W. Misner and D. H. Sharp, Phys. Rev. \textbf{136} (1964) B571.

\bibitem{R22} H. Kodama, Prog. Theor. Phys. \textbf{63} (1980) 1217.

\bibitem{R24} P. F. Gonzalez-Diaz and C. L. Siguenza, Nucl. Phys. B \textbf{697} (2004) 363.

\bibitem{R24A} M. Cataldo, P. Labrana, S. del Campo, J. Crisostomo and P. Salgado, Phys. Rev. D \textbf{78} (2008) 104006.

\bibitem{R25A} H. B. Callen, John Wiley Sons, New York, NY, USA, 1985.

\bibitem{R25B} M. S. Ma and R. Zhao, Phys. Lett. B \textbf{751} (2015) 278-283.

\bibitem{R25C} S. T. Hong and S. W. Kim, Mod. Phys. Lett. A \textbf{21} (2006) 789.

\bibitem{R26} V. Faraoni, Phys. Rev. D \textbf{84} (2011) 024003.

\bibitem{R27} K. Martel and E. Poisson, Am. J. Phys. \textbf{69} (2001) 476.
\end{thebibliography}
\end{document}